\documentclass[american]{elsarticle}
\usepackage[T1]{fontenc}
\usepackage[latin9]{inputenc}
\usepackage{float}
\usepackage{amsmath}
\usepackage{amssymb}
\usepackage{graphicx}

\makeatletter

\providecommand{\tabularnewline}{\\}

\makeatother

\usepackage{babel}
\begin{document}

\begin{frontmatter}{}

\title{Geometric calibration of the SND detector electromagnetic calorimeter}

\author[binp,nsu]{A.A. Korol}

\author[binp]{N.A. Melnikova\corref{meln}}

\ead{n.a.melnikova@inp.nsk.su}

\cortext[meln]{Corresponding author}

\address[binp]{Budker Institute of Nuclear Physics, Novosibirsk, 630090, Russia}

\address[nsu]{Novosibirsk State University, Novosibirsk, 630090, Russia}
\begin{abstract}
This paper presents the design, implementation and validation of the
software alignment procedure used to perform geometric calibration
of the electromagnetic calorimeter with respect to the tracking system
of the SND detector which is taking data at the VEPP-2000 $e^{+}e^{-}$
collider (BINP, Novosibirsk). This procedure is based on the mathematical
model describing the relative calorimeter position. The parameter
values are determined by minimizing a $\chi^{2}$ function using the
difference between particle directions reconstructed in these two
subdetectors for $e^{+}e^{-}\rightarrow e^{+}e^{-}$ scattering events.
The results of the calibration applied to data and MC simulation fit
the model well and give an improvement in particle reconstruction.
They are used in data reconstruction and MC simulation. 
\end{abstract}

\end{frontmatter}{}

\section{Introduction }

\begin{figure}[H]
\includegraphics[width=0.45\columnwidth]{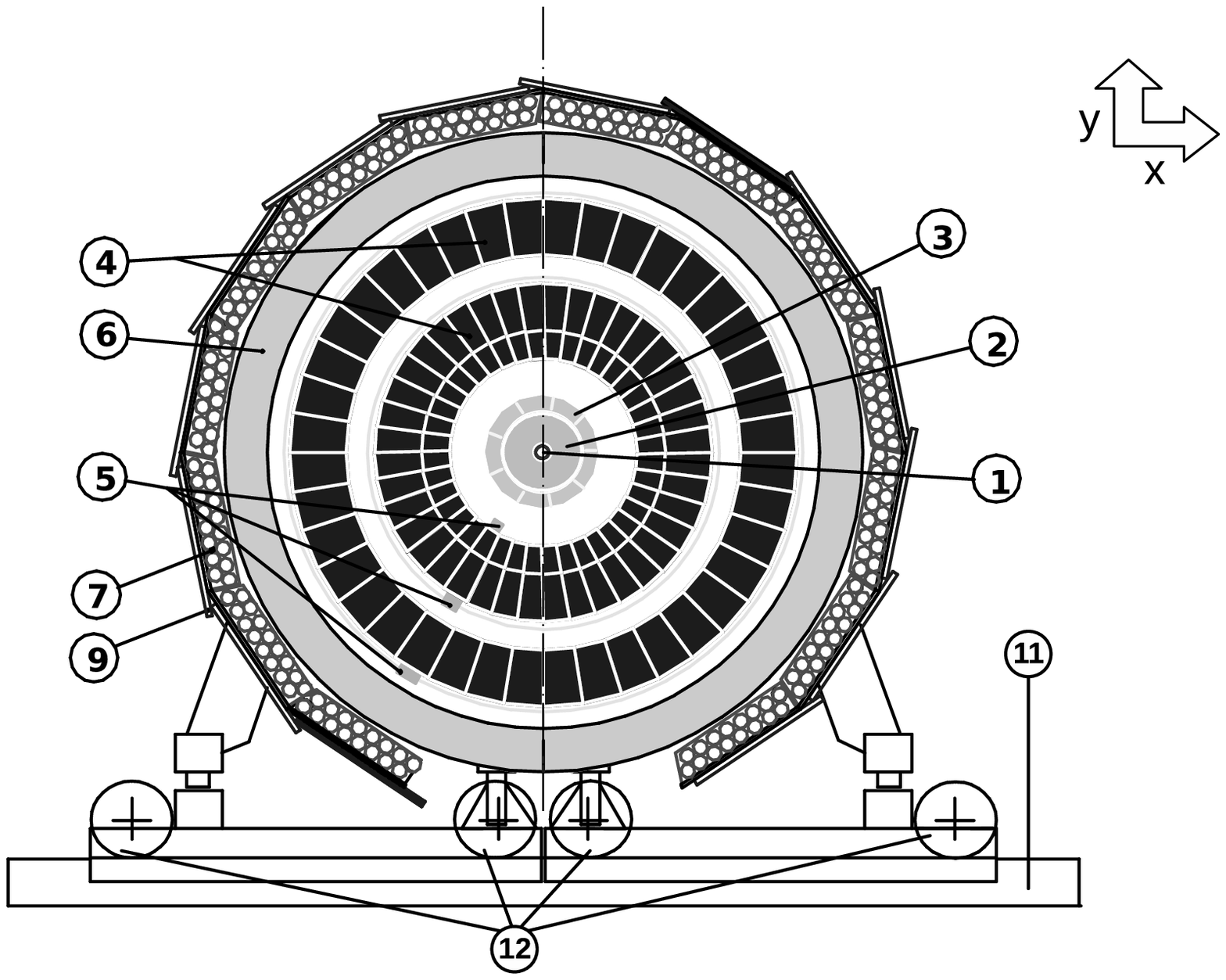}\includegraphics[width=0.45\columnwidth]{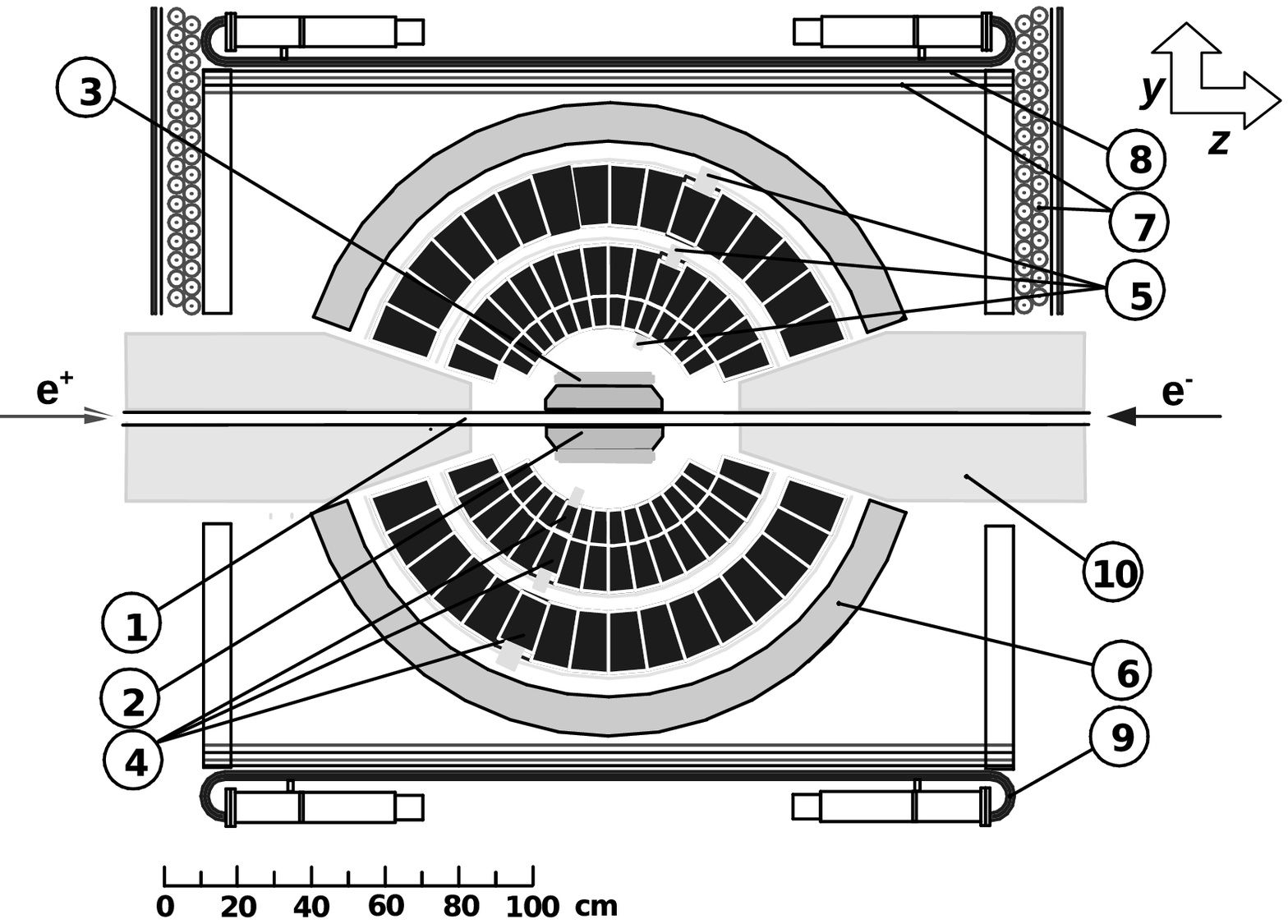}\caption{The SND scheme: 1 \textemdash{} vacuum pipe, 2 \textemdash{} tracking
system (TS), 3 \textemdash{} threshold Cherenkov counter, 4\textendash 5
\textemdash{} electromagnetic calorimeter (NaI (Tl)) (EMC), 6 \textemdash{}
iron absorber, 7\textendash 9 \textemdash{} muon detector, 10 \textemdash{}
focusing solenoids, 11 \textendash{} rails, 12 \textendash{} wheels.
\label{fig:SNDScheme:}}
\end{figure}

The Spherical Neutral Detector (SND) \cite{snd1,snd2,snd3} is used
at the electron-positron collider VEPP-2000 \cite{vepp2k} for hadronic
cross-section measurement in the center of mass energy range $0.3\div2.0$
GeV. The detector consists of several subsystems (Fig.\ref{fig:SNDScheme:})
including a spherical electromagnetic calorimeter (EMC) and a cylindrical
tracking system (TS). The detector also features a threshold Cherenkov
counters and a muon detector.

\begin{figure}
\includegraphics[width=0.95\columnwidth]{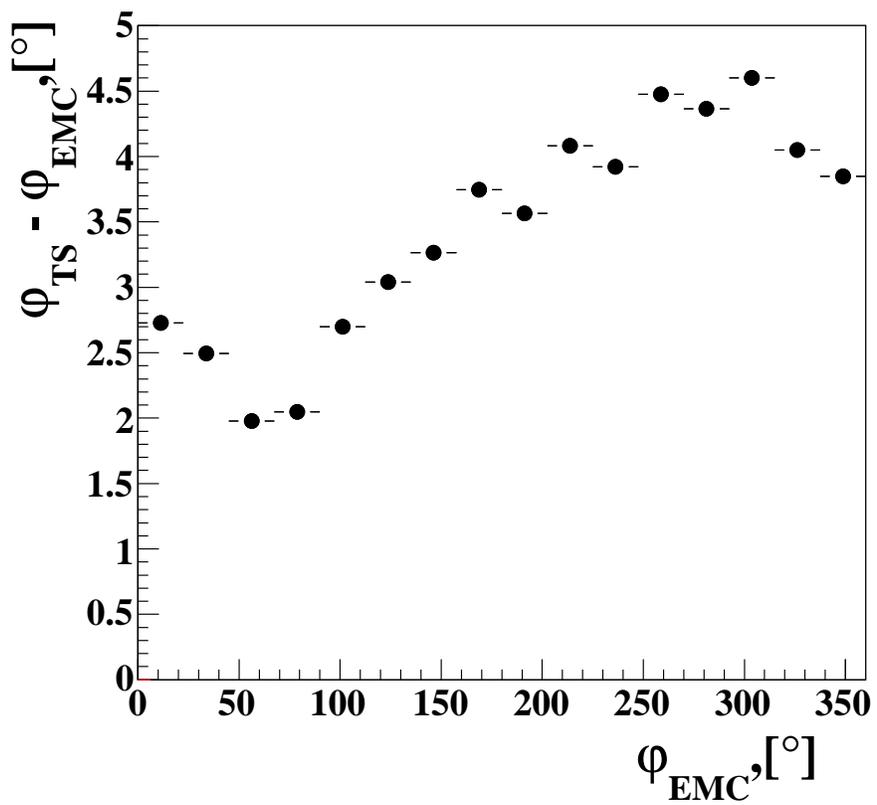}\caption{The difference between the azimuthal angles reconstructed in the TS
and in the EMC for  $e^{+}e^{-}\to e^{+}e^{-}$ events.\label{fig:Phi-problem}}
\end{figure}

The EMC is composed of 1632 NaI(TI) crystal counters arranged in 3
spherical layers. Its spherical shape covers the polar angle range
from $18^{\circ}$ to $162^{\circ}$ and provides uniform particle
detection for $95\%$ of the total solid angle. The TS is a 9-layer
drift chamber with an axial position measurement using the cathode
strips and charge division on the wires. The EMC and the TS provide
the main information for event reconstruction: the TS measures parameters
of charged particle trajectories ($z_{0},\,d_{0},\,\varphi_{TS},\,\theta_{TS}$),
while the EMC measures the energy and the angular position of electromagnetic
showers ($\varphi_{EMC},\,\theta_{EMC}$). To correctly reconstruct
physical events in the detector it is important to know precise positions
of the EMC counters relative to the TS. However data demonstrates
that the EMC is misaligned as shown at Fig.\ref{fig:Phi-problem}.

For the most interesting classes of charged particles ($\pi$- and
K-mesons) only the TS track information is relevant. On the other
hand, photons, which are in particular used to make $\pi^{0}$ and
$\eta$ meson candidates, are reconstructed only by electromagnetic
shower energy distribution in the EMC, inferring their direction from
the assumed center of the calorimeter. Consequently, even small misalignments
of the EMC and TS can be a source of kinematic inconsistency for data
analysis. Furthermore the EMC consists of two separate hemispheres
and space angles between photons could be also skewed. Fortunately,
electrons have both tracks in the TS and electromagnetic showers in
the EMC. So, we use $e^{+}e^{-}\to e^{+}e^{-}$ events to measure
the EMC position relative to the TS.

Different alignment solutions are well described in the literature
for homogeneous multicomponent detectors (e.g. vertex detectors).
Techniques based on track residual minimization are usually used for
tracking detectors. These methods minimize distance between the measured
hits and the corresponding reconstructed hits determining the detector
element position in space. Such algorithms are being used for the
ATLAS, CMS, LHCb detectors (LHC collider), for the STAR detector (RHIC
collider), for the BaBar detector (PEP-II collider), for the H1 detector
(HERA collider) etc. \cite{alignCERN}. However, there is a lack of
published results for heterogeneous systems (like EMC vs TS). While
such calibration studies were definitely conducted for the ATLAS \cite{ATLASalign},
KEDR \cite{KEDR}, CMD-3\cite{CMD3barr}, BaBar \cite{BabarEMC} experiments
(as we know from private communications) they were not openly documented.
The complication here is different sets of reconstructed parameters
for subdetectors: a 5-parameter helix (or a 4-parameter line in space
in our case) for various tracking systems versus 2-parameter shower
direction from the center reconstructed in the EMC calorimeters.

\section{Calorimeter position parametrization}

The EMC counters are rigidly fixed to two supporting aluminium spheres:
the first is for the two inner layers of counters, while the second
is for the third layer. In addition, the EMC is divided into two halves,
so that when disassembling the SND these two hemispheres can be easily
moved in the horizontal direction orthogonal to the detector axis
using rails and wheels (as illustrated on Fig.\ref{fig:SNDScheme:}
labels 11 and 12). This model of assembling and disassembling the
EMC allows to achieve a position accuracy of $3\div5\,\mathrm{\text{mm}}$.
To improve it the software geometric calibration procedure is performed
using  $e^{+}e^{-}\rightarrow e^{+}e^{-}$ events reconstructed both
in the EMC and in the TS. 

The mathematical model of the EMC sensitive element (counter) positions
relative to the TS considers the EMC halves as rigid bodies. Counter
positions take into account a possible EMC misalignment as a whole
relative to the TS and relative misalignments of the EMC halves. As
the counters are rigidly fixed to the supporting spheres with the
position accuracy of not worse than $0.03$ cm individual displacement
of each counter is small to be ignored and not considered here. 

\begin{figure}
\includegraphics[width=0.95\columnwidth]{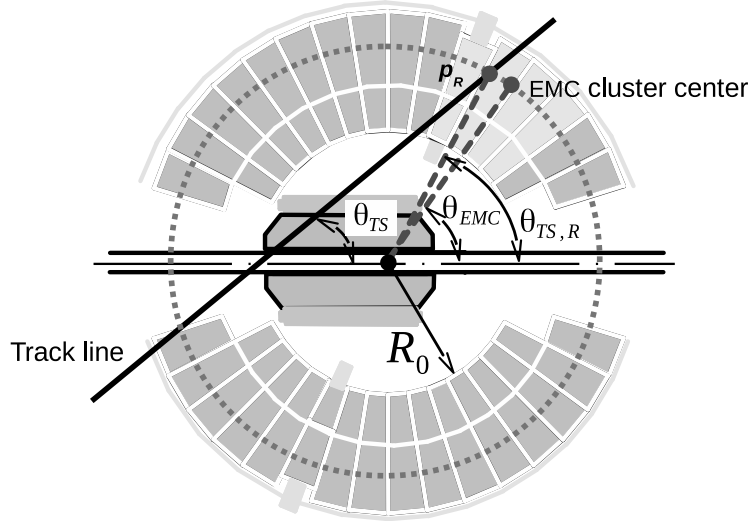}

\caption{Effective shower position in the EMC reference system. The notations
is described in the main text. \label{fig:Effective-shower-position}}
\end{figure}

We measure the EMC misalignment relative to the TS using angular differences
$\sin\left(\varphi_{TS,R}-\varphi_{EMC}\right)\,$ and $\,\theta_{TS,R}-\theta_{EMC}$,
where $\,\varphi_{TS,\,R}\,$ and $\,\theta_{TS,\,R}$ represent the
expectation for the EMC cluster direction from the TS track. They
are calculated as the polar coordinates of the point $\boldsymbol{p_{R}}$
(Fig.\ref{fig:Effective-shower-position}), in which the track crosses
a sphere with the radius 
\[
R=R_{0}+X_{0}\,\left(x_{max}-\frac{x_{pre}}{\sin\theta_{TS}}\right),
\]
where $R_{0}$ is the EMC inner radius, $X_{0}=2.56\,$cm is the radiation
length for NaI, $E_{c}=13.16\,$MeV is its critical energy, $x_{max}=\ln\frac{E}{E_{c}}-0.5$
is the maximum of the longitudinal electromagnetic shower distribution
\cite{x0Formula} , $x_{pre}\thickapprox0.13$ is an estimation of
the thickness of material between the interaction point and the EMC
($0.04$ in the TS \cite{x0Aul} , and $0.09$ in the Cherenkov counters
\cite{xoCount}). 

Unfortunately, the maximum of the energy deposition in the EMC is
located in the first two layers, making it impossible to estimate
the  misalignment of the third layer. 

The right handed TS coordinate system is used as a reference coordinate
system with center at the TS center and the Z axis is directed along
the TS axis. The Y axis is directed vertically upwards, while the
X axis is directed to the collider ring center. The EMC coordinate
system without misalignment is the same. 

\begin{figure}
\includegraphics[width=0.45\columnwidth]{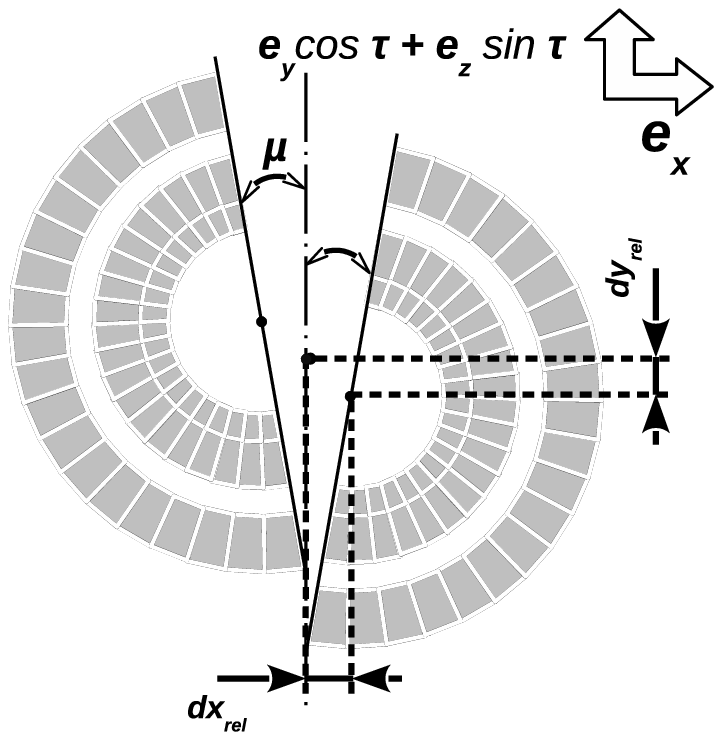}\includegraphics[width=0.45\columnwidth]{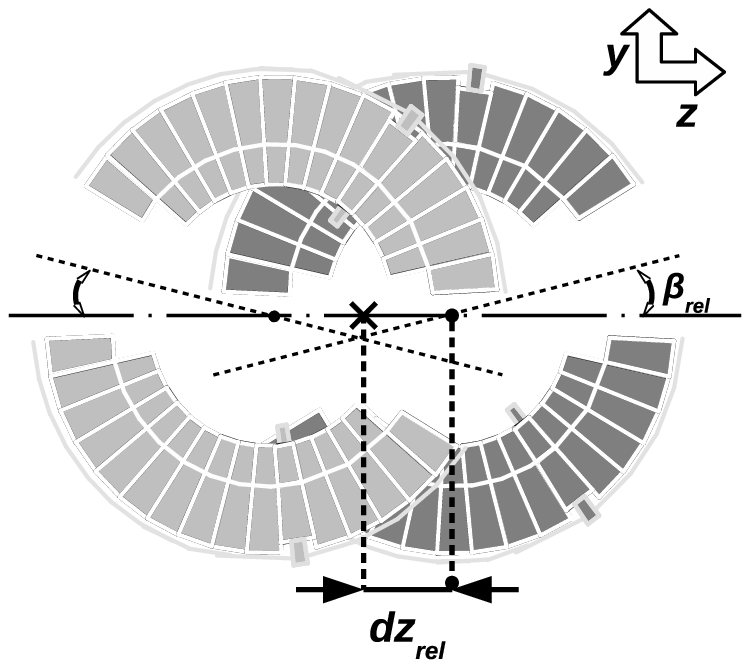}\caption{The relative parameters in the EMC reference system: left - the hemisphere
separation and the relative shift along the X axis ($dx\!_{rel}$);
right - the relative rotation around the X axis ($\beta\!_{rel}$)
and the relative shift along the Z axis ($dz\!_{rel}$).\label{fig:razval}}
\end{figure}

To describe the EMC position relative to the TS, two sets of parameters
are used. The first set describes the rotation and the shift of the
EMC as a whole:
\begin{enumerate}
\item $\alpha$ is the angle of the EMC rotation around the TS Z axis $\left(0^{\circ}<\alpha<360\textdegree\right)$;
\item $\beta$ is the EMC tilting angle with respect to the TS Z axis ($0^{\circ}<\beta<90^{\circ}$);
\item $\psi$ is a direction of the $\beta$ tilting ($0^{\circ}<\psi<180^{\circ}$);
\item $dx,\,dy,\,dz$ are the EMC shifts in corresponding directions relative
to the TS.
\end{enumerate}
The second set of parameters describes rotations and shifts of the
hemispheres relative to the EMC (Fig.\ref{fig:razval}):
\begin{enumerate}
\item $\mu$ is the half-angle of the EMC hemispheres separation $\left(0^{\circ}<\mu<90\textdegree\right)$; 
\item $\tau$ is the separation direction angle from the vertical direction
$\left(-180\textdegree<\tau<180\textdegree\right)$; 
\item $dx\!_{rel},\text{\ensuremath{dy\!_{rel}},\,\ensuremath{dz\!_{rel}}}$
are the hemisphere relative shifts;
\item $\beta\!_{rel}$ is the angle of the relative rotation around the
EMC X axis $\left(0^{\circ}<\beta\!_{rel}<360^{\circ}\right)$.
\end{enumerate}
These 12 parameters give a complete description of the hemisphere
position for the first two layers. The choice of parametrization reflects
our understanding of actual misalignment sources and their influence
to the reconstruction.

Now let $\mathbf{p_{0}}$ denote the point inside the aligned EMC
hemisphere. The common EMC rotation can be represented by the matrix
$\mathbf{\boldsymbol{T_{zyz}}}$ of the ZYZ rigid body rotation to
the Euler angles relative to the TS center: 
\begin{equation}
\mathbf{T_{zyz}}=\mathbf{R_{z}}(\psi)\cdot\mathbf{R_{y}}(\beta)\cdot\mathbf{R_{z}}(\alpha-\psi).
\end{equation}
Hereinafter $\mathbf{R_{\boldsymbol{a}}}(\mbox{\ensuremath{\gamma}})$
stands for the rotation around the $\boldsymbol{a}$ axis by an angle
$\gamma$. The common shift vector is described as:
\begin{equation}
\mathbf{\xi}=\begin{pmatrix}dx\\
dy\\
dz
\end{pmatrix}.
\end{equation}
The hemisphere separation can be described as a rotation around the
axis perpendicular to the separation direction ( $\tau$):
\begin{eqnarray*}
\mathbf{T_{\mu}} & = & \mathbf{R_{n}}(s\mu),\\
\mathbf{n} & = & \begin{pmatrix}0\\
-\sin\tau\\
\;\cos\tau
\end{pmatrix},
\end{eqnarray*}
where $s=sign(\cos\varphi_{EMC})$ is the sign of the hemisphere rotation
or the shift. Similarly, the EMC hemisphere relative shift 
\begin{equation}
\mathbf{\xi\!_{rel}}=s\cdot\begin{pmatrix}dx\!_{rel}\\
dy\!_{rel}\\
dz\!_{rel}
\end{pmatrix},
\end{equation}
and the relative rotation around the X axis

\begin{equation}
\mathbf{T_{\beta_{rel}}}=\mathbf{R_{x}}(s\,\beta\!_{rel}).
\end{equation}

The resulting transformation for $\mathbf{p}_{0}$ looks like
\begin{equation}
\mathbf{p_{1}}(\mathbf{p_{0}};\boldsymbol{\eta)}=\mathbf{T_{zyz}}\cdot(\mathbf{T_{\mu}}\cdot\mathbf{T_{\beta\!_{rel}}}\cdot\mathbf{p_{0}}+\mathbf{\xi\!_{rel}})+\mathbf{\xi},\label{eq:CalCriistal}
\end{equation}
where $\mathbf{p_{1}}$ is the point of the misaligned EMC and $\boldsymbol{\eta}$
represents a set of 12 alignment parameters. As we take the hemispheres
as rigid bodies the same transformation for counter centers and cluster
positions can be used. The experimental distributions can be described
by model functions for $\sin(\varphi_{TS,R}-\varphi_{EMC})$ and $\theta_{TS,R}-\theta_{EMC}$
as
\begin{eqnarray}
f_{\varphi}(\varphi_{0},\,\theta_{0};\boldsymbol{\eta}) & = & \sin(\varphi_{1}(\mathbf{p_{0}};\boldsymbol{\eta})-\varphi_{0}),\label{eq:fPsi}\\
f_{\theta}(\varphi_{0},\,\theta_{0};\boldsymbol{\eta}) & = & \theta_{1}(\mathbf{p_{0}};\boldsymbol{\eta})-\theta_{0},\nonumber 
\end{eqnarray}
where $\varphi_{0},\,\theta_{0}$ are the $\mathbf{p_{0}}$ azimuthal
and polar angles, $\varphi_{1},\,\theta_{1}$ are the $\mathbf{p_{1}}$
azimuthal and polar angles. In data, the $\mathbf{p_{1}}$ corresponds
to the $\mathbf{p_{R}}$ estimation from reconstructed TS data as
described above.

For better convergence a number of optimizations is done during minimization.
Parameter $\psi$ is periodic, its value poorly defined at small $\beta$.
Therefore we replace the $(\psi,\,\beta)$ pair with the Cartesian
parameters $(\beta_{1},\,\beta_{2})$:
\begin{eqnarray}
\beta_{1} & = & \sin\beta\,\cos\psi\label{eq:betaD}\\
\beta_{2} & = & \sin\beta\,\sin\psi\nonumber 
\end{eqnarray}

An alternative parametrization is used also instead of the mixed angles
and distance $(\tau,\,\mu,\,dx_{rel})$ parameters set. We use the
distance only parameters $(dp_{1},dp_{2},dp_{3})$: 
\begin{eqnarray*}
dx\!_{rel} & = & \frac{dp_{1}+dp_{2}+dp_{3}}{6},\\
\tau & = & \arctan\frac{2\,t}{dp_{1}-2\,dx\!_{rel}},\\
\mu & = & \arcsin\frac{\sqrt{4\,t^{2}+(dp_{1}-2\,dx\!_{rel})^{2}}}{2\,R_{1}},
\end{eqnarray*}
where $t=\frac{dp_{3}-dp_{2}}{2\sqrt{3}}$. These parameters correspond
to the distances between the hemispheres measured at the outer radius
of the first supporting sphere ($R_{1}$) at the angles $0^{\circ}$,
$120^{\circ}$, $240^{\circ}$ from the vertical, respectively. These
parameters have the same order of magnitude and type (units) and were
measured directly early from deformation of pieces of some plastic
substance inserted between the hemispheres. Also the negative values
of these parameters' values indicates possible calibration or reconstruction
errors which should be dealt before the calibration usage. Their non-negativeness
means that the volumes of hemispheres do not overlap, it is important
for reconstruction and MC simulation.

\section{Procedure of retrieving alignment parameter values \label{sec:Proc}}

The procedure of retrieving alignment parameter values includes several
steps. First of all,  $e^{+}e^{-}\rightarrow e^{+}e^{-}$ events are
selected using the following criteria:
\begin{enumerate}
\item no reconstructed photons and exactly two charged tracks, both with
the shift along beam $\left|z\right|<10\,cm$ and the impact parameter
$\left|\rho\right|<1\,cm$;
\item the energy deposition in the EMC for each particle $0.8\,E_{beam}<E_{1,2}<1.1\,E_{beam},(i=1,2)$;
\item acollinearity in the TS in the azimuth direction $\left|180\textdegree-\left|\varphi_{1}-\varphi_{2}\right|\right|<10\textdegree$;
\item muon veto signal absence.
\end{enumerate}
The second condition effectively suppresses influence of not operating
EMC counters. 

The next step is to fill the 2D profiles of $\sin\left(\varphi_{TS,R}-\varphi_{EMC}\right)\,$
and $\,\theta_{TS,R}-\theta_{EMC}$ depending of the azimuthal angle
range from $0\textdegree$ to $360\textdegree$ and the polar angle
range from $45\textdegree$ to $135\textdegree$ using data events
reconstructed in the EMC and in the TS. The size of the polar angle
bin is set to the counter angle size, while the size of the azimuthal
angle bin is chosen to fit a whole number of the TS cell halves and
the EMC counter halves. This is done to minimize effects from local
non-uniformity in reconstructed angles (\cite{ivanchenko}). The total
number of bins is 160 ($16\,\varphi_{EMC}\times10\,\theta_{EMC}$).
Using finer granularity for binned fit or using unbinned fit would
require to take into account exact shape of the local non-uniformity
and would unnecessary complicate the procedure. 

The fit function is constructed as a sum of the squared differences
between the measured angle residuals and the expected ones parametrized
with the model functions $f_{\varphi},\,f_{\theta}$ (Eq.\ref{eq:fPsi})
scaled by the measurement error:

\begin{eqnarray}
\chi^{2}(\boldsymbol{\eta}) & = & \underset{i}{\sum}\frac{\left(\sin(\varphi_{TS,R}-\varphi_{EMC})_{i}-f_{\varphi}(\varphi_{EMC_{i}},\,\theta_{EMC_{i}};\boldsymbol{\eta})\right)^{2}}{\sigma_{\varphi_{i}}^{2}}\nonumber \\
 & + & \underset{i}{\sum}\frac{\left(\left(\theta_{TS,R}-\theta_{EMC}\right)_{i}-f_{\theta}(\varphi_{EMC_{i}},\,\theta_{EMC_{i}};\boldsymbol{\eta})\right)^{2}}{\sigma_{\theta_{i}}^{2}},
\end{eqnarray}
where the subscript index $i$ runs over all the spatial 160 bins
and denotes averaging over all the selected events in the current
bin, $\sigma_{\varphi_{i}}$ and $\sigma_{\theta_{i}}$ are the measurement
errors. Each error includes both a statistical and systematic component:
\[
\sigma^{2}=\sigma_{stat}^{2}+\sigma_{sys}^{2}.
\]
 The systematic uncertainty is attributed to the non-uniformity in
the EMC spatial reconstruction and shortcoming of the model where
we ignore 3rd layer contribution. It is expected to be the same for
polar and azimuth direction and estimated from the data to be $0.2^{\textdegree}$.
The fit function doesn't include the correlation term between azimuthal
and polar parts because they were measured using independent detector
elements. The minimization is performed using the TMinuit class from
the ROOT framework \cite{ROOT}. The fit process includes the following
stages of the $\chi^{2}$ function minimization. At the first stage
all parameters are set to be zero and all of them except $\alpha$
are fixed. At the next stages the rest of the parameters become free
in the following order: ($dx,dy$); ($dz,\beta_{1},\beta_{2}$); ($dp_{1},dp_{2},dp_{3}$);
($dy_{rel},dz_{rel}$); ($\beta_{rel}$). 

\begin{figure}
\includegraphics[width=0.45\columnwidth]{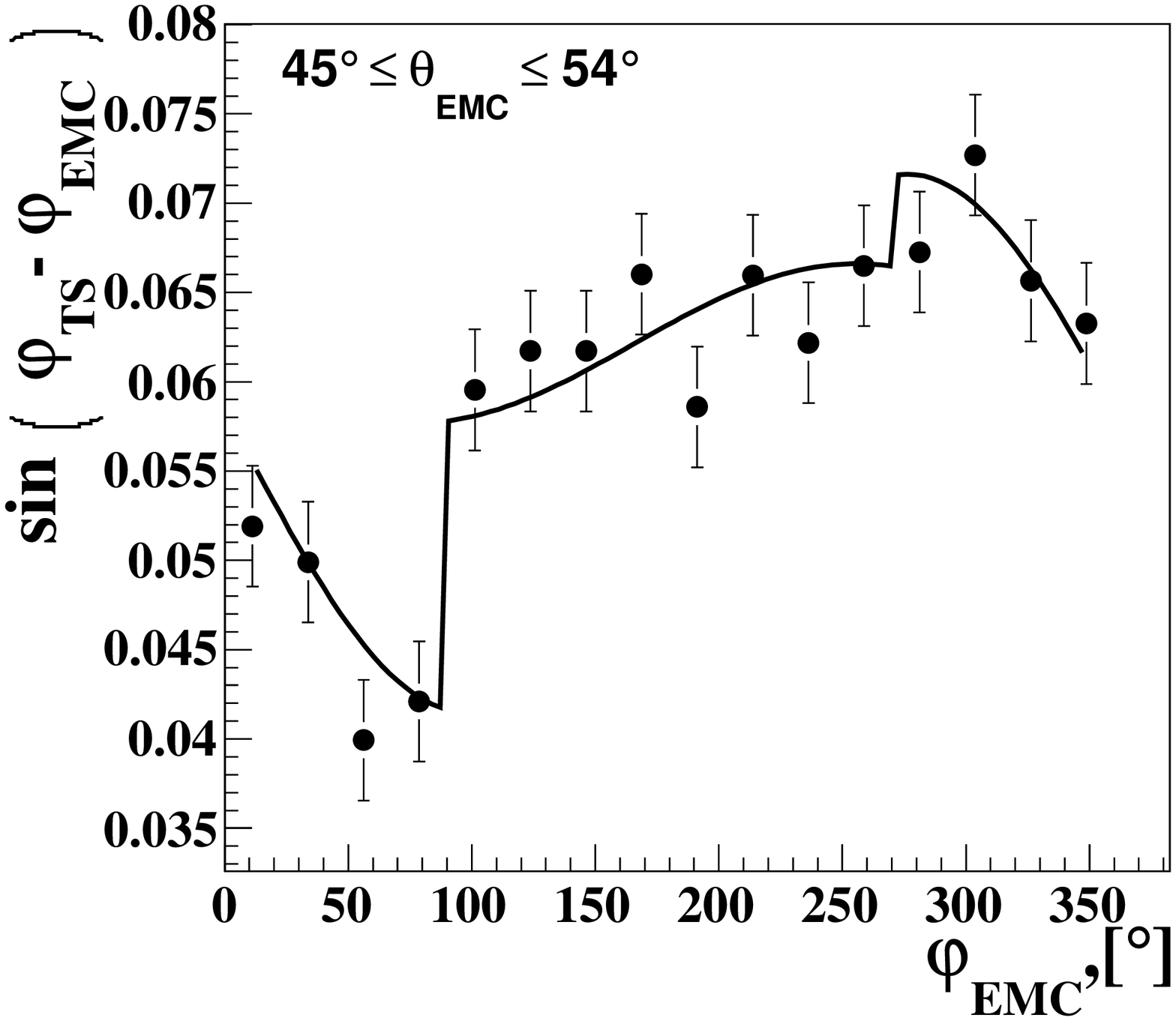}\includegraphics[width=0.45\columnwidth]{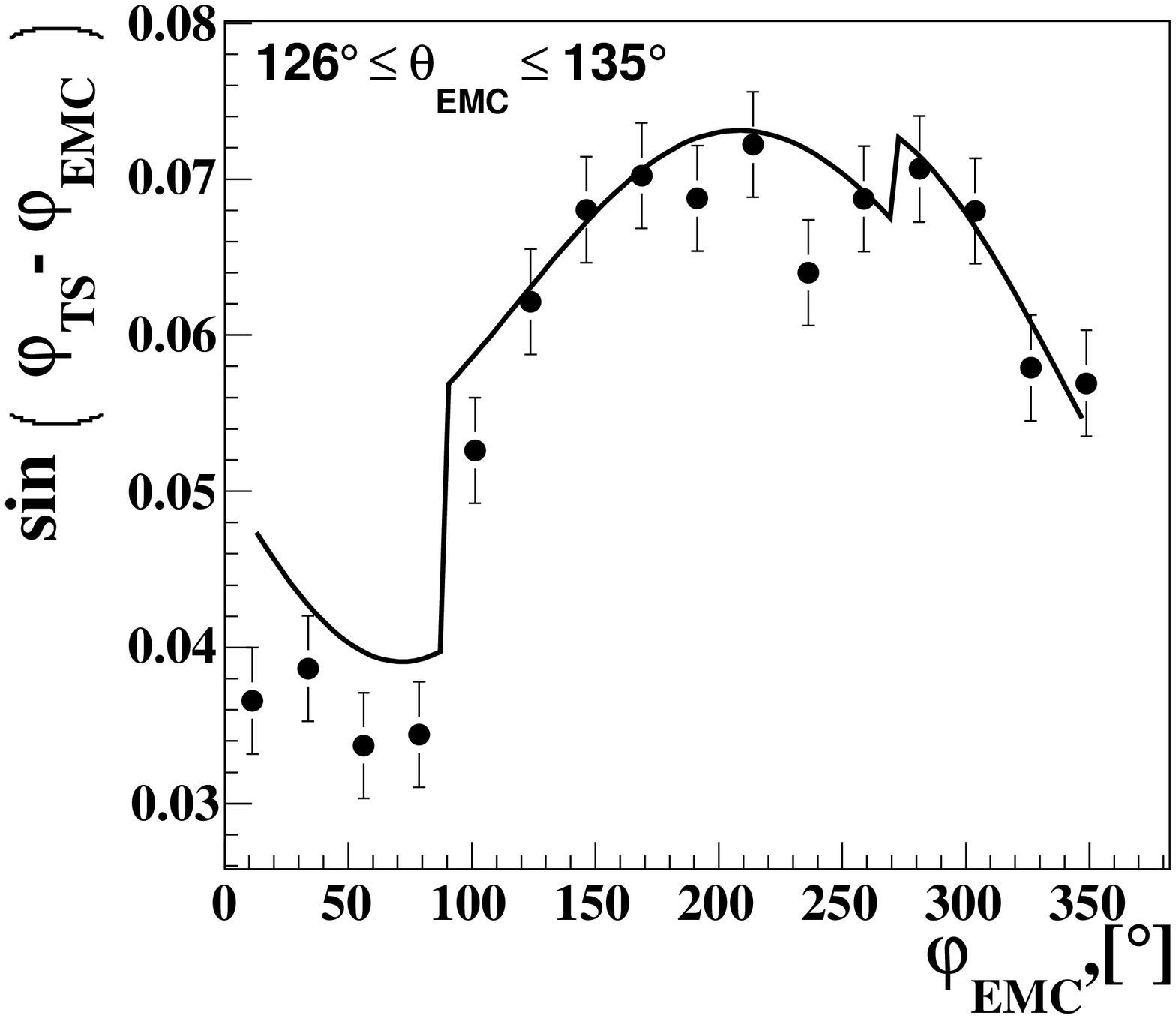}

\includegraphics[width=0.45\columnwidth]{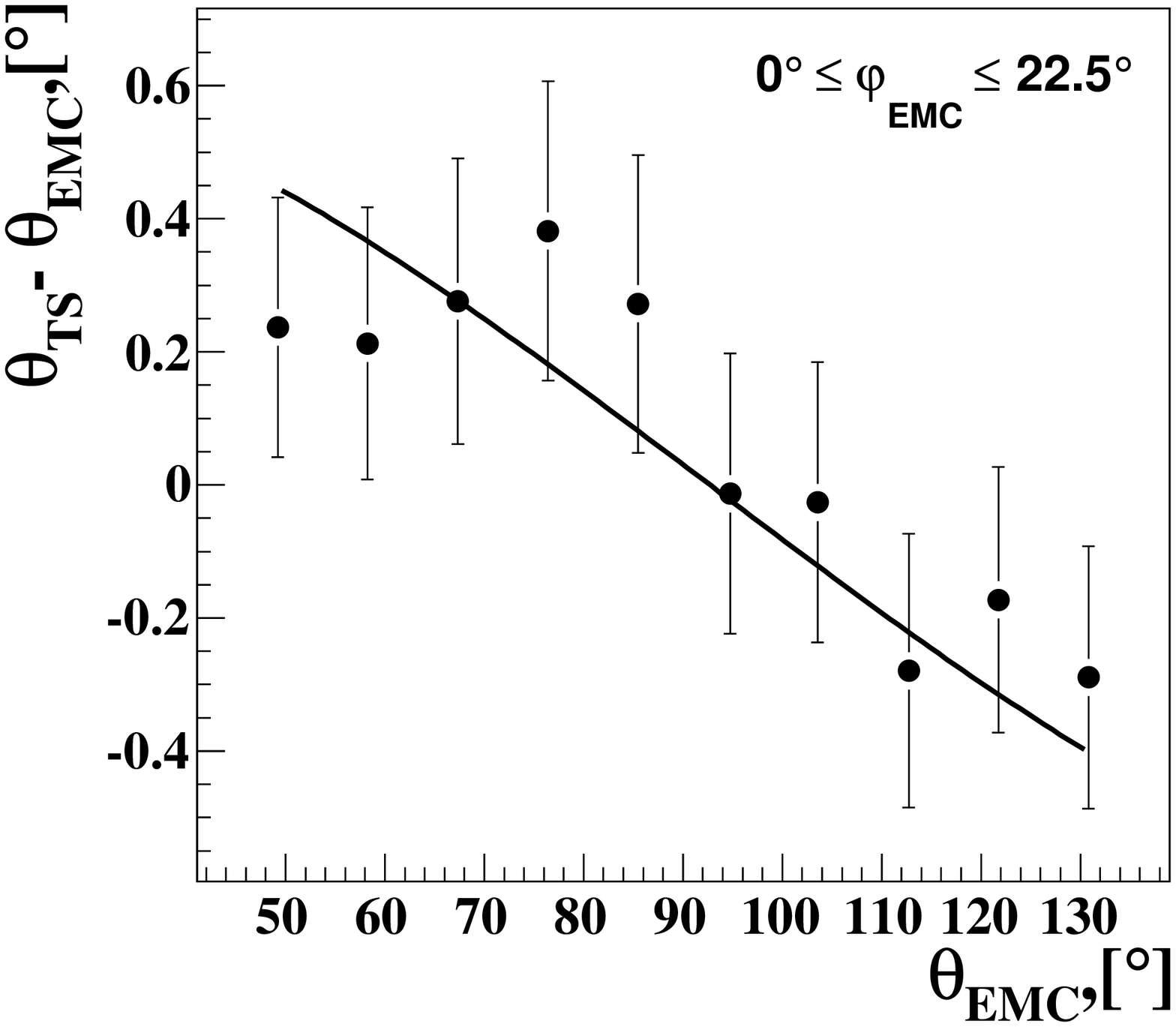}\includegraphics[width=0.45\columnwidth]{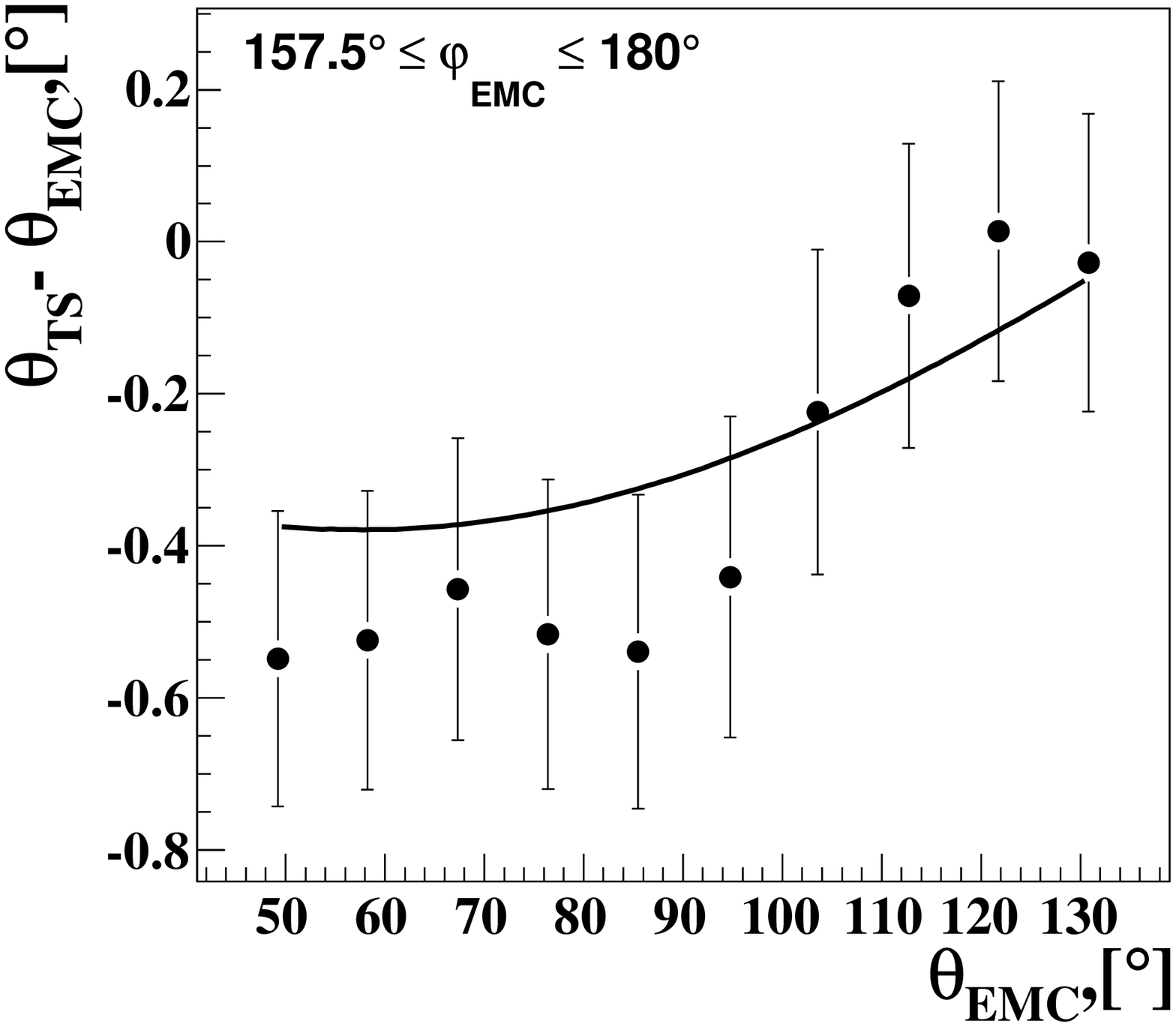}

\caption{The results of the fit to data at $E_{beam}=612.5$ MeV. The solid
lines represent the fitted functions. Points with errors show the
data distribution. The errors include the statistical and systematic
uncertainties. \label{fig: minimizationResults}}
\end{figure}

The results of a typical fit are shown in Fig.\ref{fig: minimizationResults}.
The discontinuities of the fitted curve is explained by changing the
sign of the relative translation and the relative rotation for EMC
counters close to the vertical direction. 

\section{Application of the obtained parameters }

The obtained alignment parameter values are used both in event reconstruction
and in MC simulation. Data processing is performed with the SND framework
\cite{SUMO}. Being the main part of offline SND software this framework
supports reconstruction, analysis and MC simulation.

The role of the geometric database in the framework belongs to an
object of the GeoEmcDBase class from the framework GeoDesc package.
Methods of this class transform an index of an EMC counter into geometric
coordinates (Cartesian, cylindrical or polar), which are used then
in reconstruction. Other methods of this class perform inverse transformation.
At the beginning parameters of the aligned EMC are calculated and
then they are corrected using geometric calibration results. Further
reconstruction uses these corrected EMC element coordinates. 

During reconstruction the transformation inverse to Eq.\ref{eq:CalCriistal}
is used for transforming a point of the misaligned EMC to a point
of the aligned EMC: 
\begin{equation}
\boldsymbol{p_{0}}=\mathbf{T}_{\mu}^{T}\cdot\mathbf{T}_{\beta_{rel}}^{T}\cdot(\mathbf{T_{zyz}^{T}}\cdot(\mathbf{p_{1}}-\mathbf{\mathbf{\xi}})-\mathbf{\xi\,_{rel}}).\label{eq:calCrystTrans}
\end{equation}

The alignment parameter values are determined only for the first two
EMC layers, so Eq.\ref{eq:CalCriistal} and Eq.\ref{eq:calCrystTrans}
can be safely used for clusters only in these layers. To avoid overlap
of the two third layer hemispheres, in the case if the following condition
is true (Eq.\ref{eq:dpMUCondition}):
\begin{eqnarray}
dx\!_{rel}-R_{\boldsymbol{p_{3}}}\,\sin\mu & < & 0,\label{eq:dpMUCondition}
\end{eqnarray}
 clusters of this layer are corrected using the same formulas but
with $\mu$ substituted by $\mu_{3}$:
\begin{equation}
\mu_{3}=\arcsin\frac{dx\!_{rel}}{R_{\boldsymbol{p_{3}}}},
\end{equation}

where $R_{\boldsymbol{p_{3}}}$ is the outer radius of the second
supporting sphere.

For realistic MC simulation volumes representing the EMC hemispheres
are placed in the world volume taking into account the obtained alignment
parameter values using Eq.\ref{eq:CalCriistal}. This is performed
by applying corresponding transformation to the nested volumes. Due
to the hierarchy of the volumes there no need to place individual
counters. The third layer hemispheres are placed taking into account
the above considerations (Eq.\ref{eq:dpMUCondition}) in order to
perform correct MC simulation.

\section{Results and discussion}

The alignment algorithm is validated using MC simulation, that takes
into account the alignment parameters to reproduce the EMC displacement
effects observed in data. To improve the quality of the simulation,
non-operating EMC counters are masked. To take into account beam-generated
spurious hits in the detector in MC, special events recorded during
experiment with a random trigger are merged with the simulated events
\cite{Kardapo}.

To validate the algorithm, 100 MC simulation samples are produced
with different sets of the alignment parameters evenly distributed
over the expected range. Used statistics of $50000$ events per sample
is compatible with the statistics used for one calibration for data.
The biases and spreads of the parameters, estimated as the means and
RMSs of the differences between the values obtained from the fit and
the true MC values, are listed in Table \ref{tab:MCFitResults}. The
largest biases are observed for $\alpha$ and $dy_{rel}$ . These
parameters also have the largest negative correlation $(-0.85)$ between
them because the relative shift along the Y axis of the EMC halves
in opposite directions adds an effective bias of the rotation angle
around the Z axis. The biases do not exceed $0.2^{\circ}$ for angles
and $0.06$ cm for translations and are too small to lead to observable
shifts in reconstructed photon parameters. Therefore, we conclude
that the true and measured parameter values are in reasonable agreement. 

\begin{table}
\begin{tabular}{|c|c|c|c|c|c|}
\hline 
EMC & bias & spread & Halves & bias & spread\tabularnewline
\hline 
\hline 
$\alpha,\:^{\circ}$ & $-0.13$ & $0.02$ & $\mu,\:^{\circ}$ & $-1\,10^{-4}$ & $3\,10^{-4}$\tabularnewline
\hline 
$\beta_{1},\:^{\circ}$ & $-0.007$ & $0.023$ & $dx\!_{rel},$ cm & $-0.021$ & $0.008$\tabularnewline
\hline 
$\beta_{2},\:^{\circ}$ & $-0.006$ & $0.017$ & $dp_{1},$cm & $-0.045$ & $0.031$\tabularnewline
\hline 
$dx,$ cm & $-0.005$ & $0.006$ & $dp_{2},$cm & $-0.030$ & $0.030$\tabularnewline
\hline 
$dy,$ cm & $-0.005$ & $0.012$ & $dp_{3},$ cm & $-0.053$ & $0.036$\tabularnewline
\hline 
$dz,$ cm & $0.012$ & $0.021$ & $\beta\!_{rel},\:^{\circ}$ & $-1\,10^{-4}$ & $5\,10^{-4}$\tabularnewline
\hline 
 &  &  & $dy\!_{rel},$ cm & $0.052$ & $0.018$\tabularnewline
\hline 
 &  &  & $dz\!_{rel},$ cm & $-0.005$ & $0.018$\tabularnewline
\hline 
\end{tabular}\caption{The biases and spreads of the alignment parameters obtained using
MC simulation.\label{tab:MCFitResults}}
\end{table}

\begin{figure}
\includegraphics[width=0.45\columnwidth]{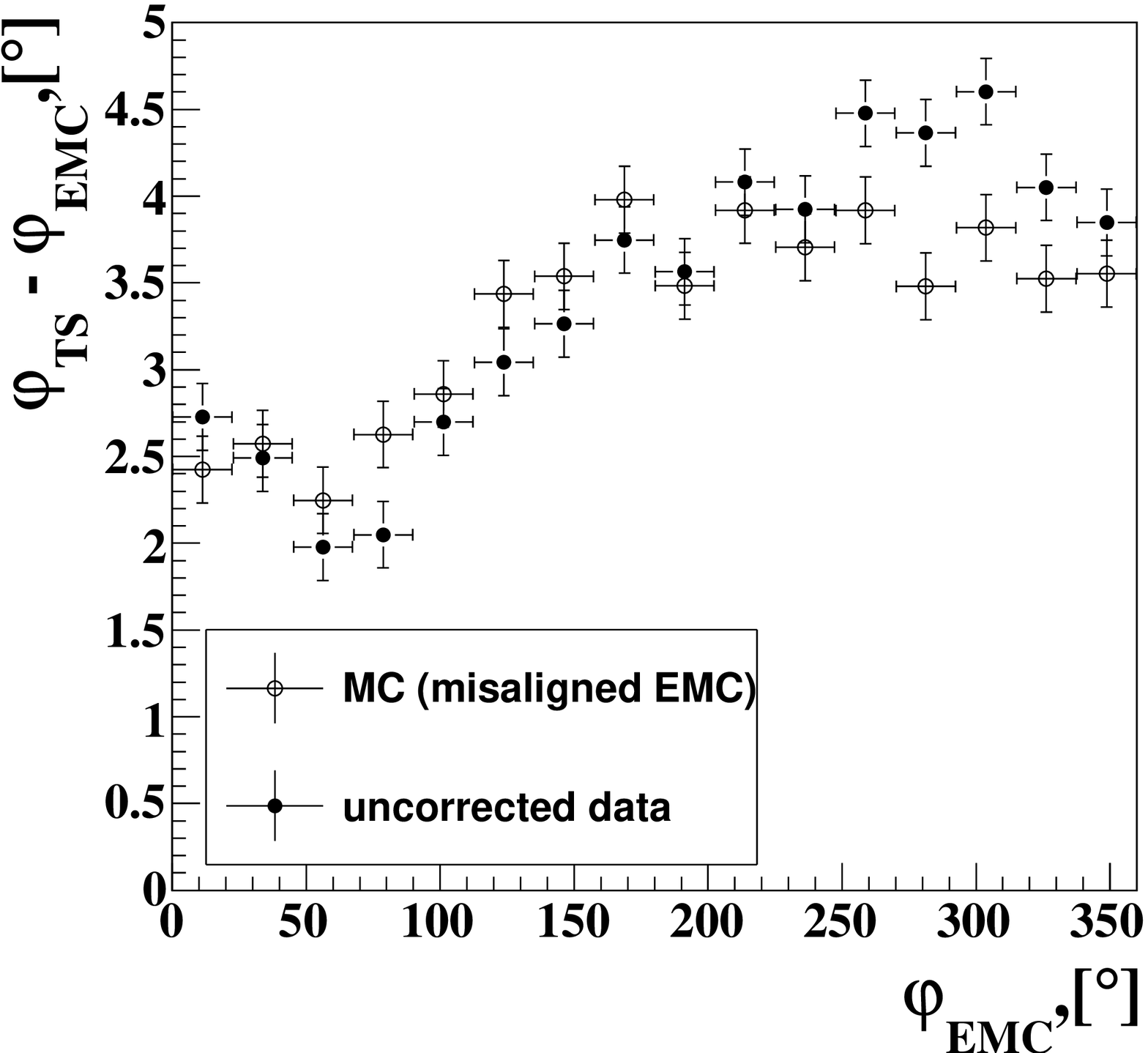}\includegraphics[width=0.45\columnwidth]{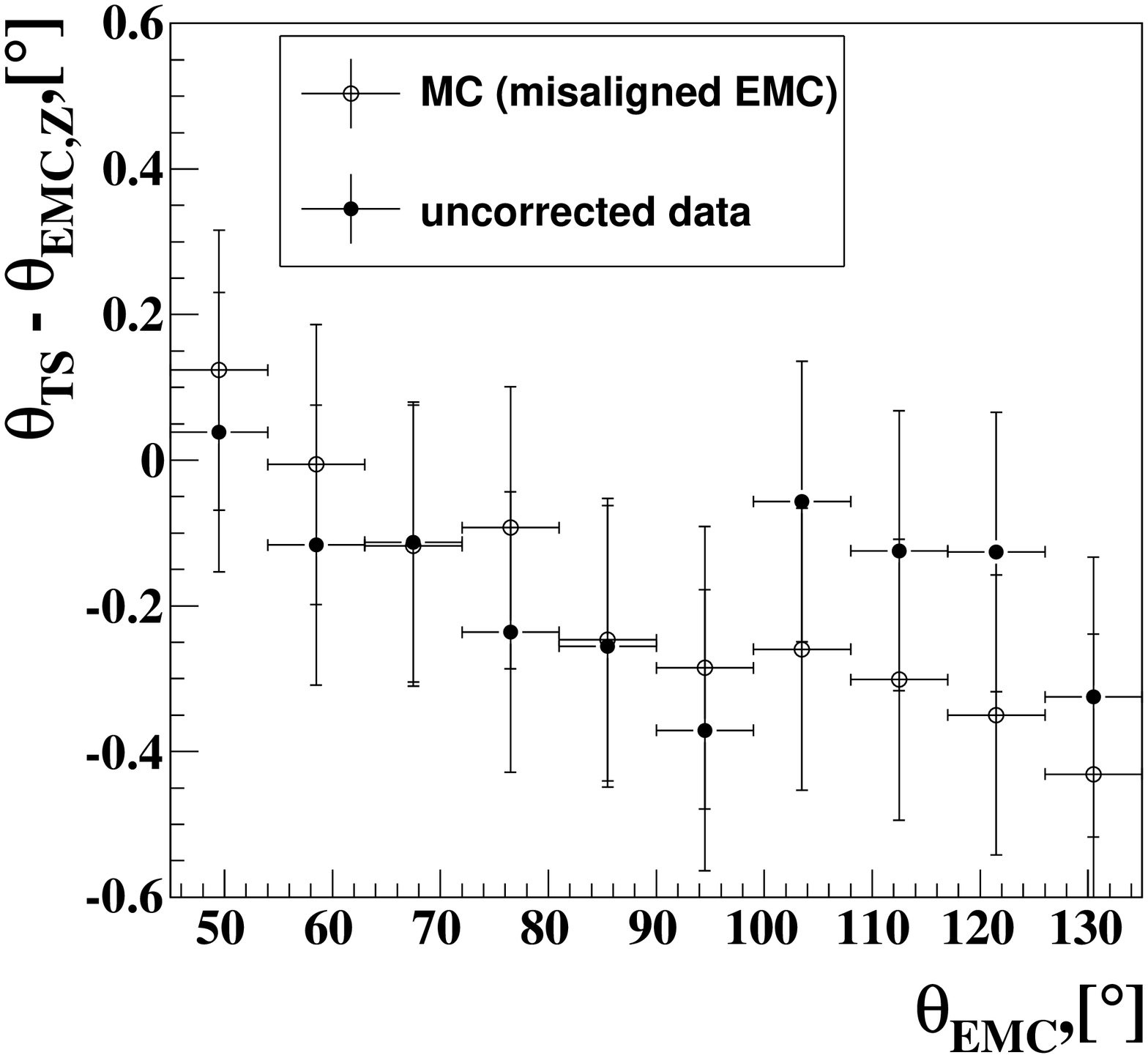}\caption{Comparison of the difference between the angles reconstructed in the
EMC and in the TS for $e^{+}e^{-}\rightarrow e^{+}e^{-}$ events in
uncorrected data (filled circle) and realistic MC simulation (empty
circle) at $E_{beam}=612.5\,$MeV. In calculation of $\theta_{EMC,z}$
the shift of the event vertex along the Z axis is taken into account.
The errors include the statistical and systematic uncertainties. \label{fig:mcVSData}}
\end{figure}

Comparison of the described above realistic MC simulation with the
uncorrected data, shown in Fig.\ref{fig:mcVSData}, demonstrates satisfactory
agreement. 

\begin{figure}
\includegraphics[width=0.45\columnwidth]{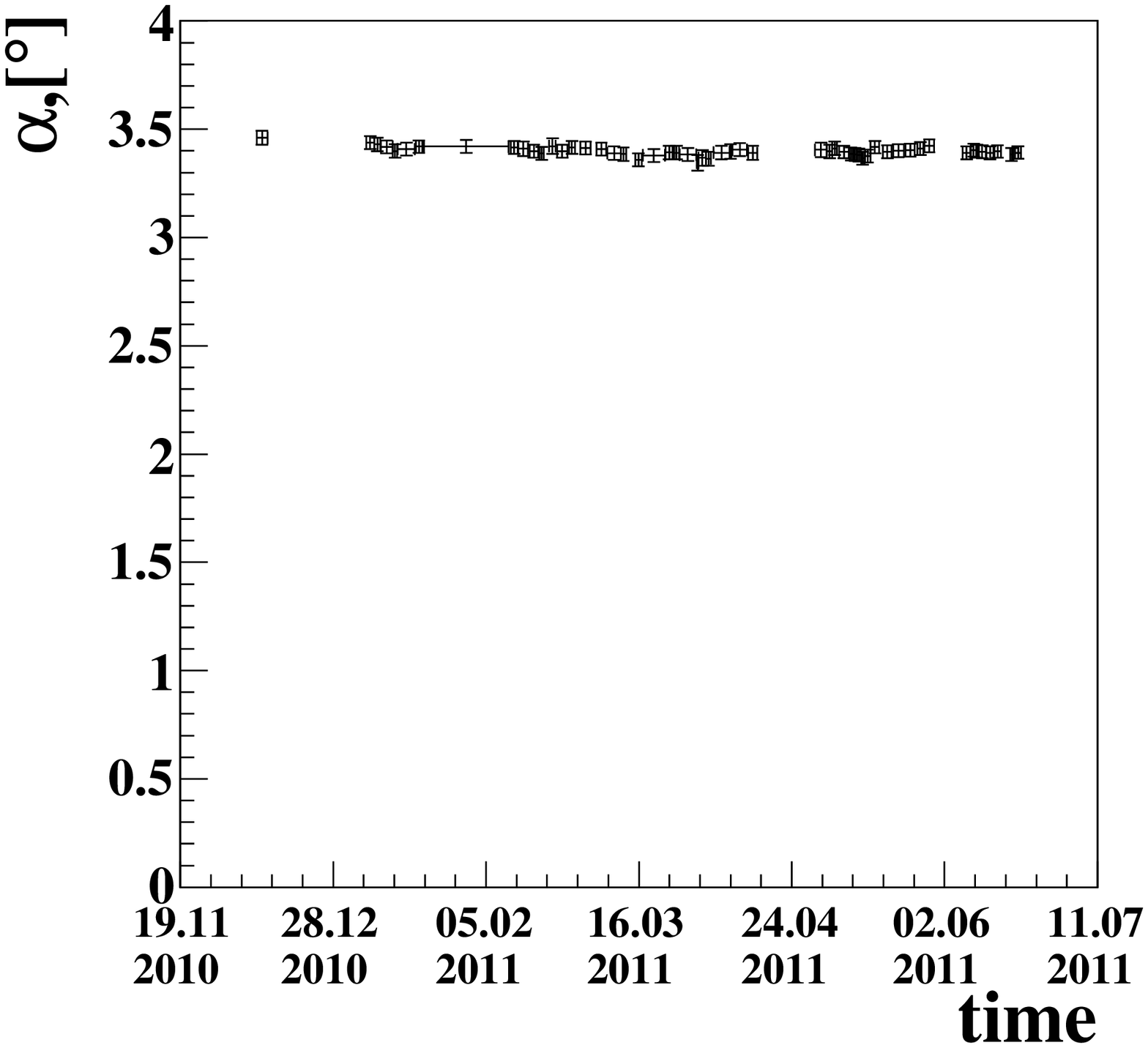}\includegraphics[width=0.45\columnwidth]{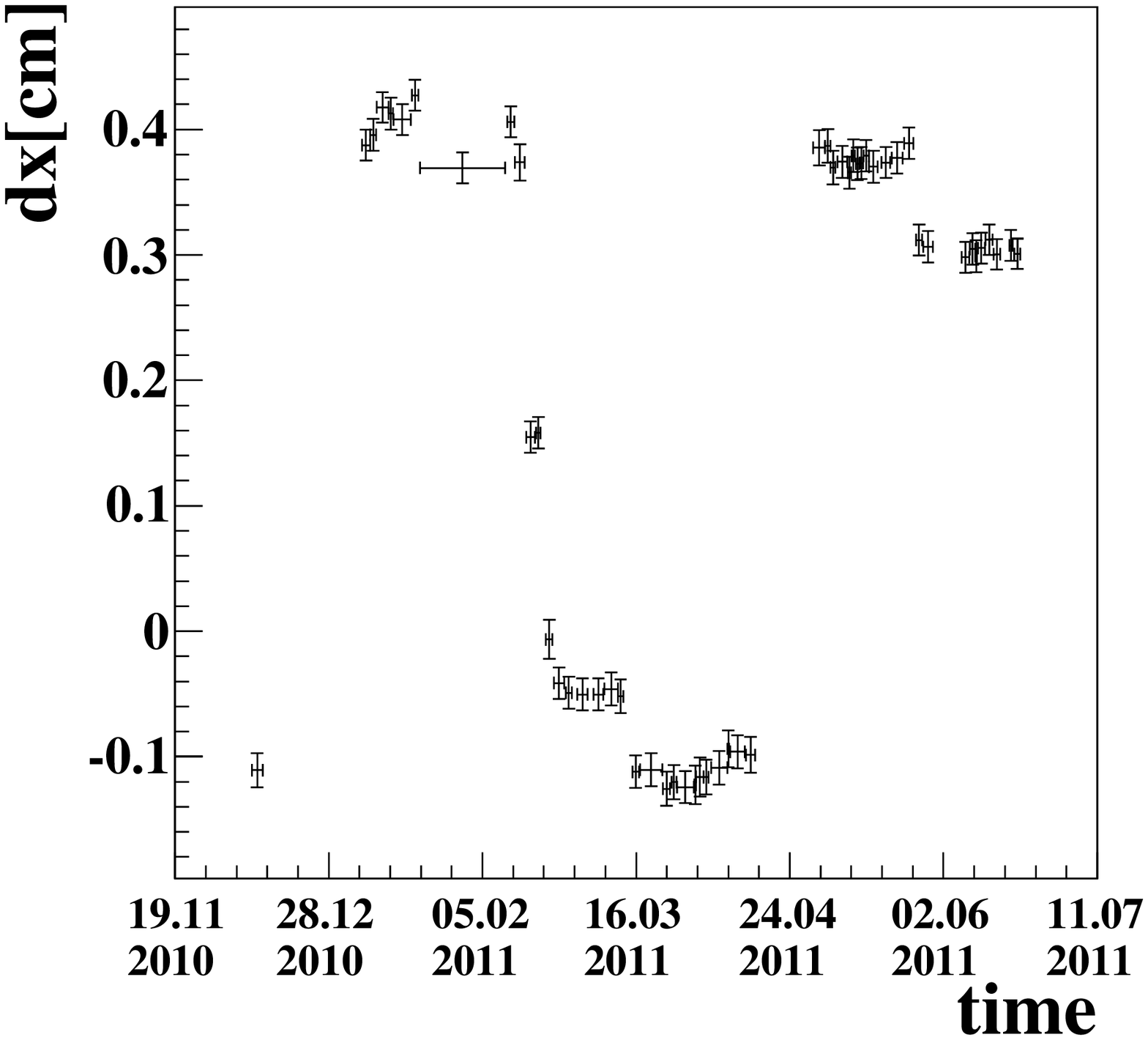}

\includegraphics[width=0.45\columnwidth]{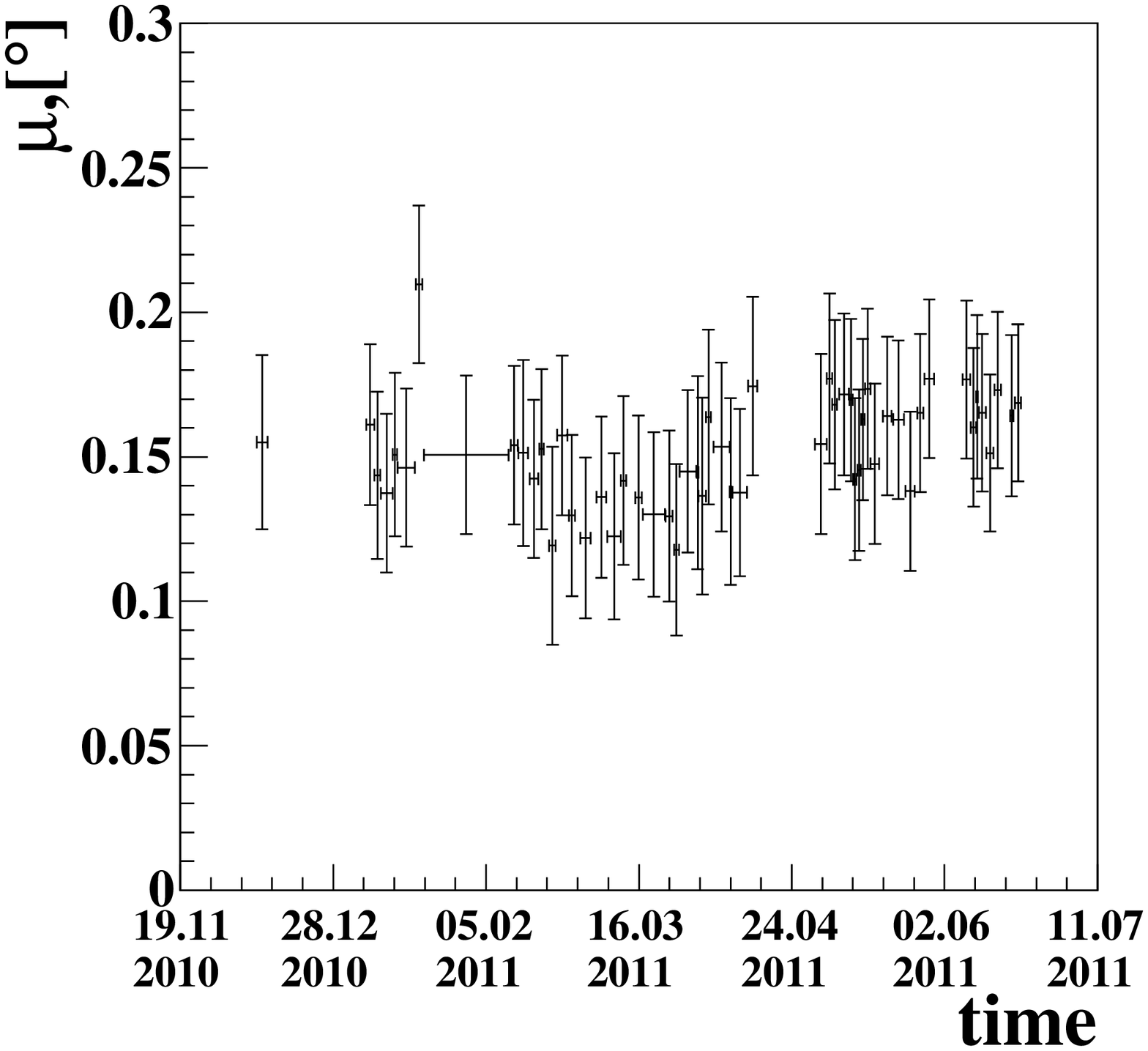}\includegraphics[width=0.45\columnwidth]{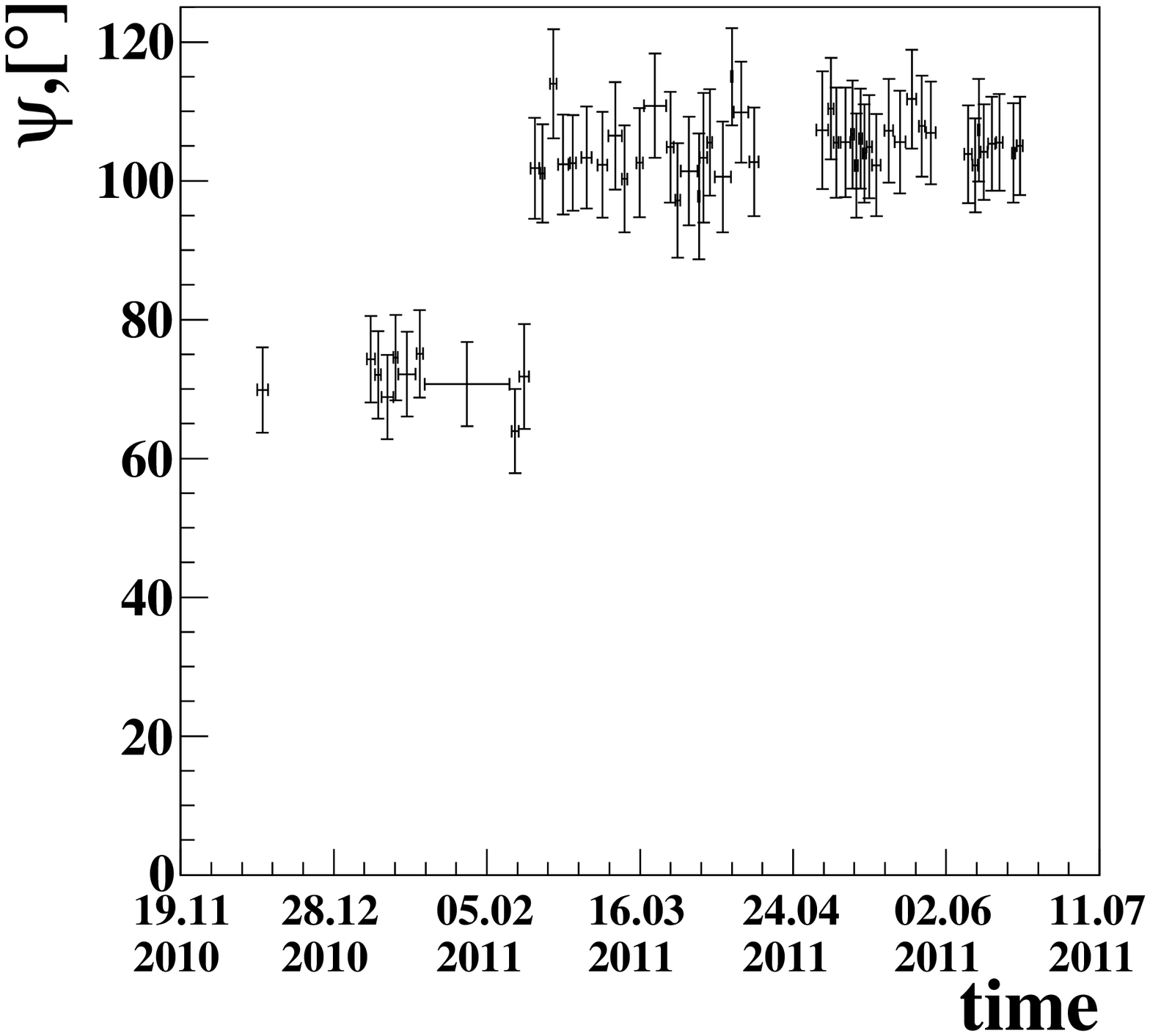}\caption{The time evolution of the values of the parameters $\alpha$, $dx$,
$\mu$, $\psi$.\label{fig:ParamsValuesMHAD2011}}
\end{figure}

The procedure described in Sec.\ref{sec:Proc} is applied to about
$2000$ runs during 2010-2011 recorded with an integrated luminosity
of $25.3\,pb^{-1}$. In Fig.\ref{fig:ParamsValuesMHAD2011} the dependence
of obtained alignment parameters on time is shown. The value of $\alpha$
was stable during this period, while the value of $dx$ changed several
times due to disassembling and reassembling the detector. One of these
mechanical interruptions also changed the angle between the TS and
the EMC axes ($\psi$). We do not see steady changes in the EMC geometry
parameters. This supports our assumption on these changes origin.
It is also possible to use these graphs for quality control: steady
changes in parameter values indicate incorrect event reconstruction
in the TS or in the EMC. The obtained relative shift along the X axis
and the relative rotation around the Z axis are close to measured
with instrumental methods.

\begin{figure}
\includegraphics[width=0.95\columnwidth]{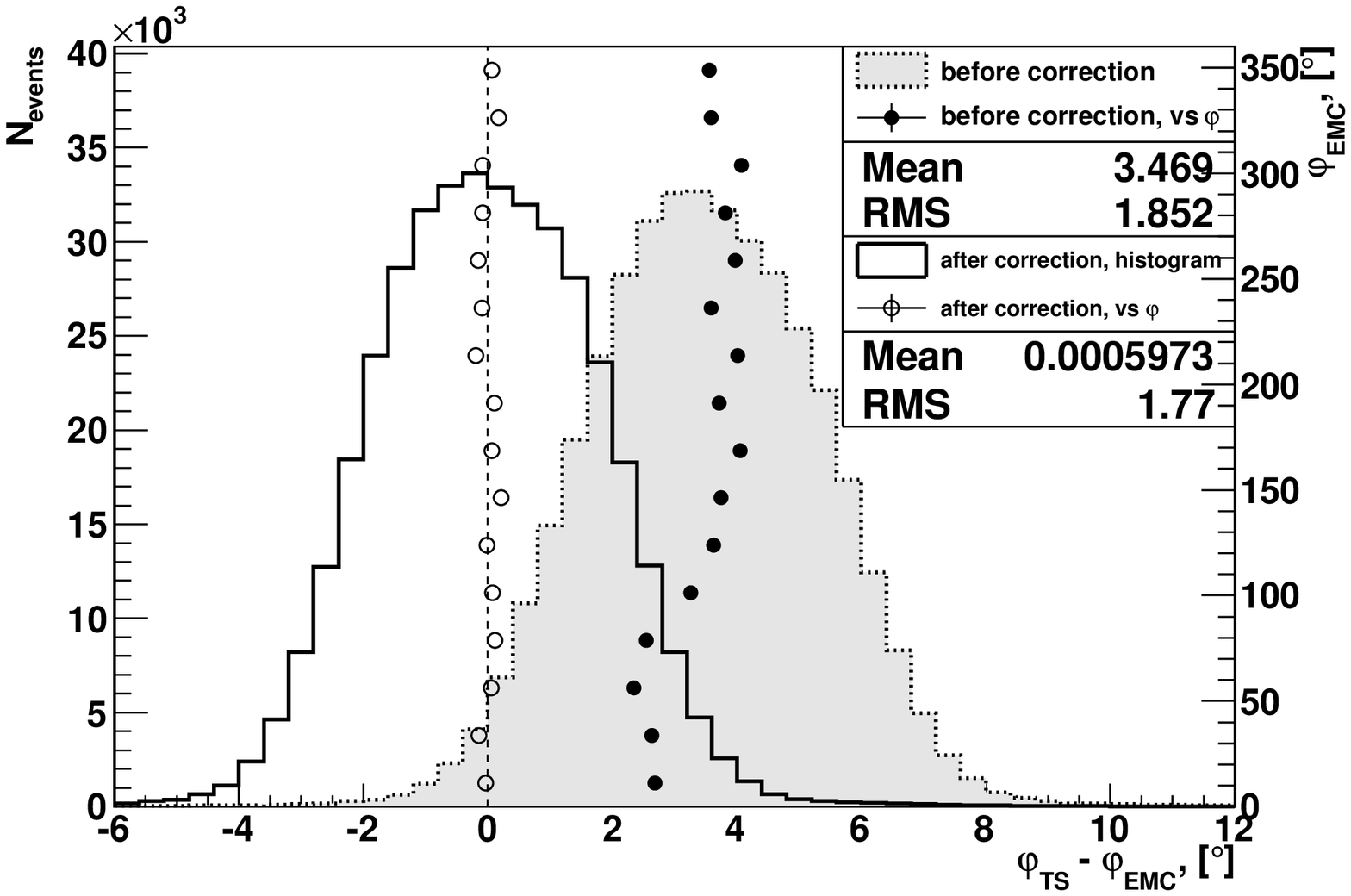}

\includegraphics[width=0.95\columnwidth]{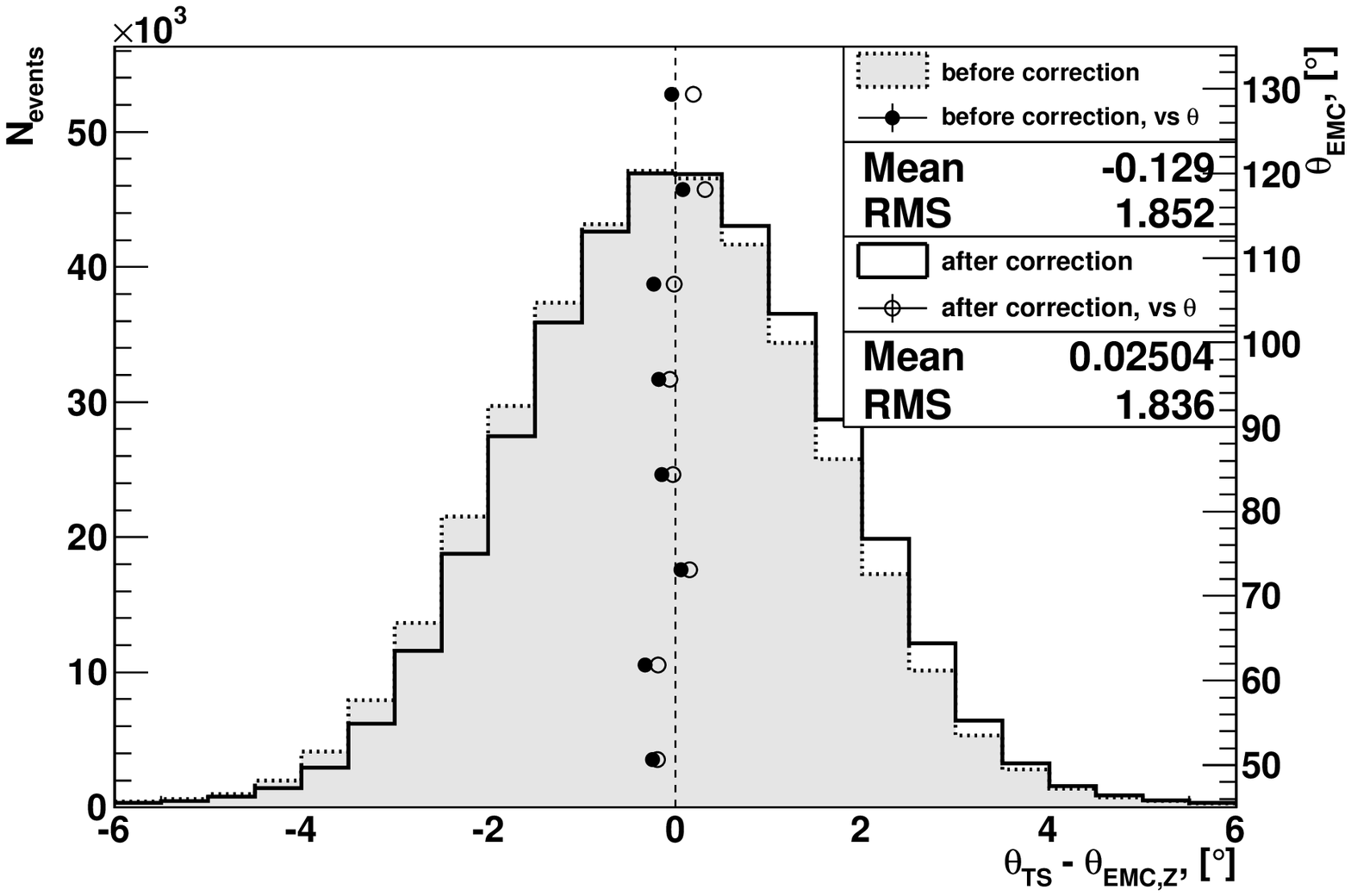}\caption{Results of the geometric correction for $e^{+}e^{-}\rightarrow e^{+}e^{-}$
data at $E_{beam}=612.5$ MeV. The left scale and filled and opened
histograms represent the difference between angles reconstructed in
the EMC and in the TS before and after the corrections respectively.
The right scale and filled and opened circles represent the same difference
as functions of corresponding EMC angles before and after the corrections
respectively. In calculation of $\theta_{EMC,z}$ the shift of the
event vertex along the Z axis is taken into account. \label{fig:AlignmentResultsBaBa}}
\end{figure}

\begin{figure}
\includegraphics[width=0.95\columnwidth]{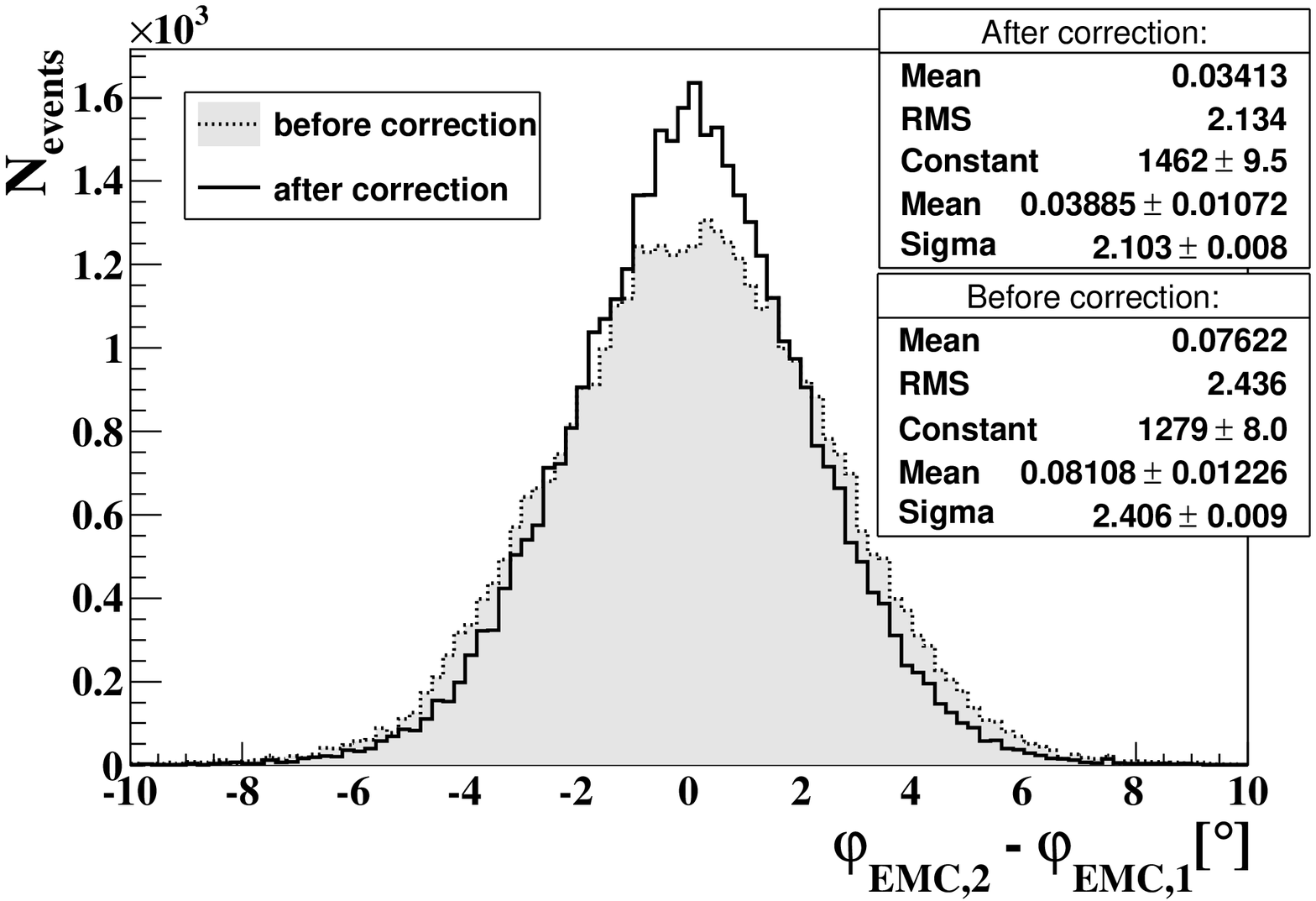}\caption{The result of the geometric correction for $e^{+}e^{-}\rightarrow2\gamma$
events at $E_{beam}=612.5$ MeV. Filled and opened histograms represent
the difference between angles reconstructed in the EMC and in the
TS before and after the corrections respectively.\label{fig:AlignmentResultsExpDeltaPhi_2g}}
\end{figure}

The results of the geometric alignment ($E_{beam}=612.5\,$MeV) for
$e^{+}e^{-}\to e^{+}e^{-}$ events are shown in Fig.\ref{fig:AlignmentResultsBaBa}.
Taking into account the relative rotation angle of the EMC halves
can eliminate its systematic contribution to the difference between
the angles of photons traversing the different EMC halves. This difference
can be observed in the acollinearity distribution for the  $e^{+}e^{-}\to2\gamma$
events (Fig.\ref{fig:AlignmentResultsExpDeltaPhi_2g}) selected with
the following conditions:
\begin{enumerate}
\item exactly two reconstructed photons and zero charged tracks;
\item the energy deposition in the EMC for each particle $E_{1,2}>0.7\,E_{beam}$;
\item difference between particle energies $|E_{1}-E_{2}|<150\,$MeV;
\item muon veto signal absence;
\item $36\textdegree<\theta_{EMC_{1,2}}<144\textdegree$.
\end{enumerate}

Figure \ref{fig:AlignmentResultsExpDeltaPhi_2g} demonstrates $16\%$
improvement in the EMC angular resolution due to the correction.

\begin{figure}
\includegraphics[width=0.95\columnwidth]{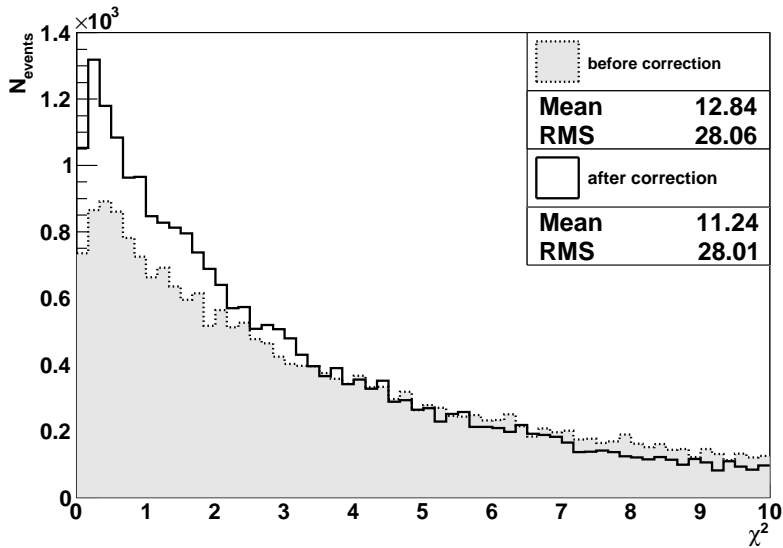}\caption{The kinematic fit quality parameter for  $e^{+}e^{-}\rightarrow\pi^{+}\pi^{-}\pi^{0}$
events. Filled and opened histograms represent the difference between
angles reconstructed in the EMC and in the TS before and after the
corrections respectively. \label{fig:pi0Xi2}}
\end{figure}

To study the influence of the geometric calibration on the reconstruction
of events containing both charged particles and photons (mixed charge).
The process $e^{+}e^{-}\rightarrow\pi^{+}\pi^{-}\pi^{0}$ is used.
The candidate events from the energy range near the $\phi(1020)$
resonance are selected using the following conditions:
\begin{enumerate}
\item exactly two charged particles, two neutral particles;
\item $36^{\circ}<\theta_{chr\,1,2}<144^{\circ}$;
\item $\left|180\textdegree-\left|\varphi_{chr\,1}-\varphi_{chr\,2}\right|\right|>10^{\circ}$;
\item $\arccos\left(n_{chr\,1}\cdot n_{chr\,2}\right)<147\textdegree;$
\item the energy deposition form photons $E_{N\,1}+E_{N\,2}>100\,MeV$.
\end{enumerate}
Then a kinematic fit is performed with the requirements of energy
and momentum conservation and without the $\pi^{0}$ mass constraint.
The fit uses photon angles and energies and only angles for charged
particles. It should be noted that our detector does not measure momenta
of charged particles. The corrections lead to significant improvement
of the fit quality (Fig.\ref{fig:pi0Xi2}). It should be noted that
because non-Gaussian distribution of measured parameters fit quality
parameter not necessary follows the $\chi^{2}$ distribution.

\section{Conclusions}

We have developed and implemented the geometric calibration procedure
for the SND detector electromagnetic calorimeter. The procedure is
validated using MC simulation. It is shown that most of the alignment
parameters values stay stable during the data taking. Some of them
slightly changes due to disassembling and reassembling the detector.

The procedure allows us to reduce the difference between the angles
measured in the TS and the EMC. The common bias in the azimuthal angle
about $3\textdegree$ and its irregularity of about $1\textdegree$
are removed. The angular correction allows us to improve the kinematic
fit quality for mixed charge events (the $\chi^{2}$ mean value decreases
by $10-15\,\%$ ).

The results of the geometric calibration are used in the data analysis
and  MC simulation \cite{etaprime2015,lkl2015,snd2016}. We think
that the ideas implemented in the calibration procedure can be useful
in other experiments with heterogeneous detectors as well.

\section{Acknowledgments }

This work was supported by the Russian Science Foundation {[}grant
number 14-50-00080{]}.

\bibliographystyle{elsarticle-num}
\bibliography{article_en_src}

\end{document}